\documentclass[lettersize,journal]{IEEEtran}
\usepackage{amsmath,amsfonts}
\usepackage{amssymb}
\usepackage{algorithmic}
\usepackage{algorithm}
\usepackage{array}
\usepackage{textcomp}
\usepackage{float}
\usepackage{subfigure}
\usepackage{multirow}
\usepackage{url}
\usepackage{verbatim}
\usepackage{graphicx}
\usepackage{cite}
\usepackage{xcolor}
\begin{document}

\title{Feasibility Study Regarding Self-sustainable Reconfigurable Intelligent Surfaces}

\author{Zhenyu~Li,
        Ozan~Alp~Topal,
        {\"O}zlem Tu\u{g}fe Demir,
        Emil Bj{\"o}rnson,
        Cicek Cavdar
\thanks{Z. Li, O. A. Topal, E. Björnson, and  C. Cavdar are with the School of Electrical Engineering and Computer Science, KTH Royal Institute of Technology, Stockholm, Sweden (e-mail: \{zhenyuli, oatopal, emilbjo, cavdar\}@kth.se).} 
\thanks{Ö. T. Demir is  with the Department of Electrical and Electronics Engineering, Bilkent University, Ankara, Türkiye, (e-mail: ozlemtugfedemir@bilkent.edu.tr).}  
\thanks{This study is conducted under the Eureka Celtic Project RAI-6Green: Robust and AI Native 6G Green Mobile Networks (ID: C2023/1-9) and partly supported by Swedish Wireless Innovations Center: SweWIN (2023-00572), both funded by Swedish Innovation Agency Vinnova.}
}



\maketitle

\begin{abstract}
    Without requiring operational costs such as cabling and powering while maintaining reconfigurable phase-shift capability, self-sustainable reconfigurable intelligent surfaces (ssRISs) can be deployed in locations inaccessible to conventional relays or base stations, offering a novel approach to enhance wireless coverage. This study assesses the feasibility of ssRIS deployment by analyzing two harvest-and-reflect (HaR) schemes: element-splitting (ES) and time-splitting (TS). We examine how element requirements scale with key system parameters, transmit power, data rate demands, and outage constraints under both line-of-sight (LOS) and non-line-of-sight (NLOS) ssRIS-to-user equipment (UE) channels. Analytical and numerical results reveal distinct feasibility characteristics. The TS scheme demonstrates better channel hardening gain, maintaining stable element requirements across varying outage margins, making it advantageous for indoor deployments with favorable harvesting conditions and moderate data rates. However, TS exhibits an element requirement that exponentially scales to harvesting difficulty and data rate. Conversely, the ES scheme shows only linear growth with harvesting difficulty, providing better feasibility under challenging outdoor scenarios. These findings establish that TS excels in benign environments, prioritizing reliability, while ES is preferable for demanding conditions requiring operational robustness.
\end{abstract}


\section{Introduction}
    \IEEEPARstart{M}{etasurfaces}, with their low-cost manufacturing and ability to manipulate wireless environments, have been extensively studied to enhance communication quality~\cite{di2020smart}. As more prototypes emerge, it becomes evident that maintaining reconfigurable phase shifts requires cabling and external power~\cite{10177872}, which limits deployment flexibility. To look for potential solutions, our previous work analyzed the trade-off between operating costs and metasurface gains by comparing reconfigurable intelligent surfaces (RISs) and static metasurfaces (SMSs) \cite{10570594}. While RIS provides adjustable phase shifts at the expense of power consumption, SMS offers fixed phase shifts without energy needs. Our findings indicate that the throughput advantages of reconfigurability can be offset by densely deploying SMSs, though this remains costly in practice. These limitations motivate the exploration of self-sustainable reconfigurable intelligent surfaces (ssRISs). As introduced in \cite{10348506}, ssRISs offer reconfigurability like RIS while harvesting energy from incoming electromagnetic waves to power their operation. This capability makes ssRIS a promising solution for cost-effective, high-throughput metasurface-assisted systems.

    A related study~\cite{10348506} investigated the impact of different harvest-and-reflect (HaR) schemes across various communication scenarios. However, the element count requirements for achieving specific quality-of-service (QoS) targets remain underexplored. As a critical design parameter, the number of elements in an ssRIS directly determines the feasibility of satisfying the QoS requirement, as beamforming gain scales quadratically with the element count, enabling stronger signal amplification, with each additional element increasing power consumption, challenging the self-sustainability requirement. Furthermore, larger element counts escalate both manufacturing costs and control complexity, thereby imposing practical limitations on implementation feasibility. Therefore, it is essential to assess how gracefully each HaR scheme adapts to varying system conditions, such as transmit power, data rate requirements, in terms of element count requirements. We define this adaptability as \textit{feasibility}. Specifically, schemes requiring fewer elements under given system constraints, or demonstrating slower growth in element requirements as conditions deteriorate, provide greater feasibility for practical ssRIS deployment.

    This letter investigates how feasible to utilize ssRIS to enhance communication. The main contributions of this letter are summarized as follows:
    \begin{itemize} \item For a single-user equipment (UE) multiple-input single-output (MISO) system, we prove that when the ssRIS-to-base station (BS) channel is line-of-sight (LOS)-dominated, maximum-ratio transmission (MRT) precoding guarantees both optimal energy harvesting and optimal signal-to-noise ratio (SNR) at the UE when there is no direct link between the BS and UE.
    
    \item Our analytical derivations together with numerical results reveal that different HaR schemes exhibit varying feasibility under different harvesting conditions and data rate requirements. The time-splitting (TS) scheme enhances ssRIS feasibility when harvesting conditions are favorable and data rate requirements are low. Additionally, the TS scheme shows better feasibility in terms of maintaining a low outage. Conversely, in more challenging scenarios with poor harvesting conditions or high data rate demands, the element-splitting (ES) scheme better maximizes ssRIS feasibility.
    \end{itemize}

\section{System Model}

\begin{figure}[tb]
        \centering
        \includegraphics[width=0.6\linewidth]{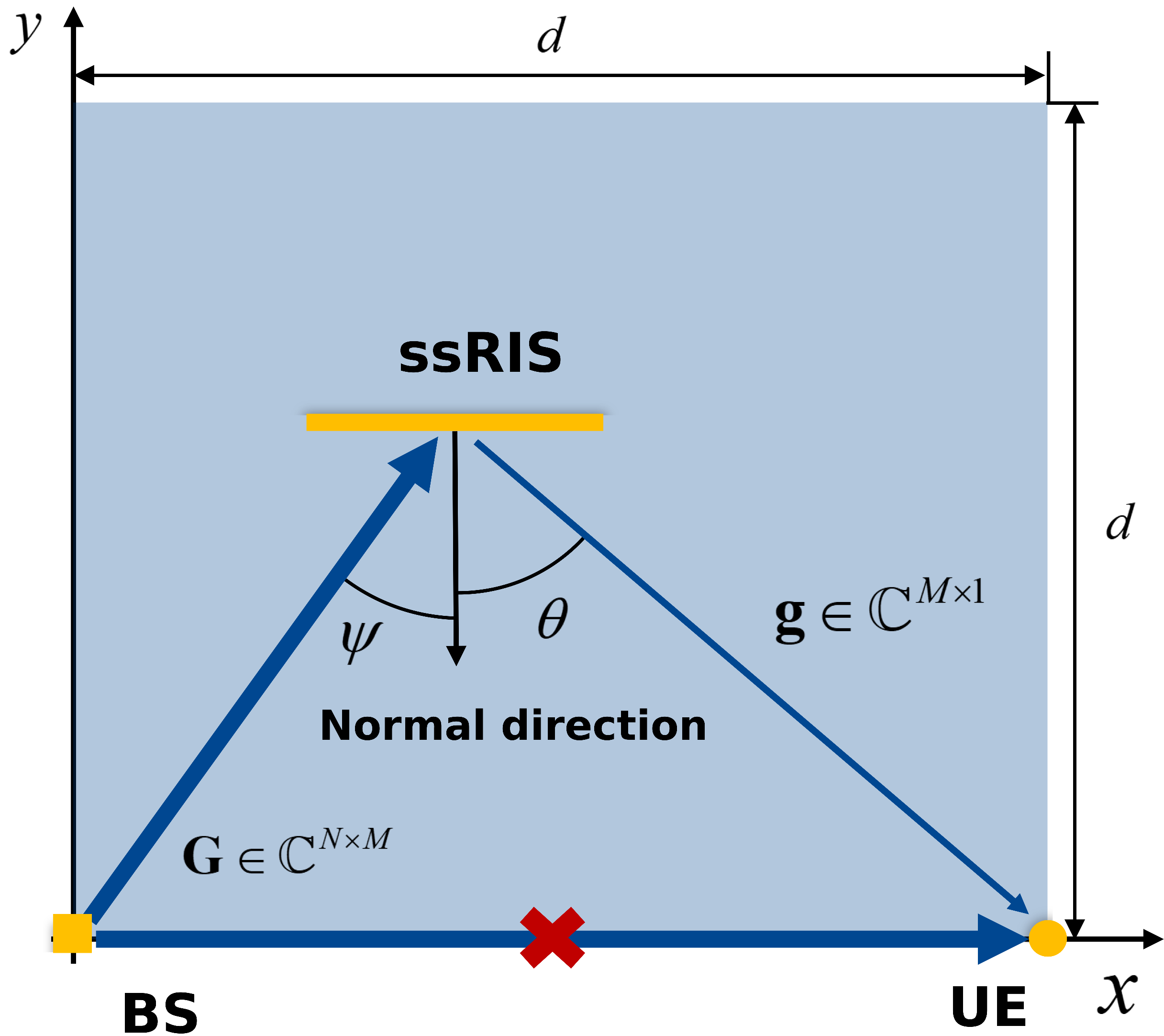}
        \caption{System model. The square and circle represent the position of the BS and the UE, respectively. }
        \label{fig:systemmodel}
    \end{figure}

   As shown in Fig.~\ref{fig:systemmodel}, our system consists of a BS, a UE, and an ssRIS positioned within a square region of side length $d$. The BS and UE are located at opposite corners. The BS is equipped with $N$ antennas in a uniform planar array (UPA), the ssRIS has $M$ elements also in a UPA, and the UE has an isotropic antenna. All devices are coplanar, with the ssRIS surface parallel to the BS plane.
   
   Since metasurface gains diminish significantly with direct BS-UE LOS links~\cite{10570594, liu2021reconfigurable}, we assume the UE has no direct link to the BS due to blockage. Additionally, for self-sustainability, the ssRIS must be deployed in the BS's LOS region to ensure sufficient energy harvesting. The BS-ssRIS channel $\mathbf{G} \in \mathbb{C}^{N \times M}$ is assumed pure LOS
    \begin{equation}
        \mathbf{G}=\sqrt{\rho(\psi,d_\text{SR})}e^{-j\varphi_\text{SR}}\cdot\mathbf{a}_N(\psi)\mathbf{a}_M^T(\psi)
    \end{equation}
    where $\varphi_{\text{SR}}$ is the reference phase shift, $d_{\text{SR}}$ is the propagation distance, and $\psi$ is the azimuth angle of the ssRIS with respect to the BS. The free-space path loss (FSPL) is $\rho(\psi, d) = \frac{\lambda^2}{(4\pi d)^2} G(\psi)$, where $\lambda$ is the wavelength and $G(\psi) = \pi\cos(\psi)$ for $\psi \in [-\pi/2, \pi/2]$ represents the ssRIS antenna gain following the cosine model~\cite{di2020smart, bjornson2024introduction}. The array response vectors $\mathbf{a}_M(\psi) \in \mathbb{C}^{M \times 1}$ and $\mathbf{a}_N(\psi) \in \mathbb{C}^{N \times 1}$ have unit-modulus entries.

    We denote the channel from ssRIS to UE by $\mathbf{g}\in\mathbb{C}^{M \times 1}$. The received signal at the UE side can be expressed as 
    \begin{equation}
        y = \boldsymbol{\phi}^H\mathbf{Hw}s+n\label{eq:receivedsignal}
    \end{equation}
    where $n$ is the additive white Gaussian noise (AWGN) with a spectral density of $N_0$. Additionally, the cascaded BS-ssRIS-UE channel is given as $\mathbf{H}\in\mathbb{C}^{M\times N}=\operatorname{diag}(\mathbf{g})\mathbf{G}^H$. The $\boldsymbol{\phi}\in\mathbb{C}^{M\times 1}$ represents the phase shifts injected by the ssRIS. Moreover, $\mathbf{w}\in\mathbb{C}^{N\times 1}$ is the unit-norm digital precoding vector, and $s$ is the data stream that satisfies $\mathbb{E}\{|s|^2\}=P$. 
    
    The power received by the ssRIS can be expressed as
    \begin{align}
        P_\text{Rc}&=P\left\|\mathbf{G}^H\mathbf{w}\right\|^2\label{eq:receivedpower}\notag\\
        &=P\rho(\psi,d_\text{SR})\left(\mathbf{w}^H\mathbf{a}_N(\psi)\mathbf{a}^T_M(\psi)\mathbf{a}^*_M(\psi)\mathbf{a}_N^H(\psi)\mathbf{w}\right)\notag\\
        &\leq PNM\rho(\psi,d_\text{SR})
    \end{align}
where the upper limit is obtained by the Cauchy-Schwarz inequality. Additionally, the power harvested from the AWGN is neglected. Notably, the precoder influences not only the signal quality received by the UE but also the self-sustainability of the ssRIS. Specifically, the inner product between $\mathbf{a}_N(\psi)$ and $\mathbf{w}$ in~\eqref{eq:receivedpower} indicates that the choice of precoding directly affects the self-sustainability condition of the ssRIS. 

\section{System design and data rate analysis}

  As analyzed earlier, the precoder must be carefully designed to maximize the self-sustainability of the ssRIS and the SNR at the UE. In this section, we first discuss the optimal design of phase shifts under the rank-1 channel condition. Then, we prove that under the assumption that the channel between ssRIS and BS is dominated by LOS component and there is no direct link between the BS and UE, MRT beamforming guarantees both maximum power harvesting at the ssRIS and maximum SNR at the UE, regardless of the specific channel realization between the UE and the ssRIS.

    \subsection{ssRIS phase-shifts design}
         For the sake of simplicity without losing generality, we assume lossless reflection. Consequently, the magnitude of each phase-shift entry, $|\phi_m| = 1$ holds for all $m$. We utilize MRT beamforming, $\mathbf{w} = \frac{(\boldsymbol{\phi}^H\mathbf{H})^H}{\|\boldsymbol{\phi}^H\mathbf{H}\|}\label{eq:mrt}$,  which maximizes the SNR of the UE. Given the LOS assumption between the BS and the ssRIS, the Gram matrix $\mathbf{V}\in\mathbb{C}^{M\times M}$, defined as $\mathbf{V}=\mathbf{HH}^H$, has rank one. Then the optimal selection of the phase shifts becomes $\boldsymbol{\phi}=\exp(j\angle\mathbf{u}_1)$, where $\mathbf{u}_1$ is the eigenvector of $\mathbf{V}$ corresponding to the only non-zero eigenvalue $\lambda_1$. Then the SNR can be given by
        \begin{equation}
            \Gamma^* = \frac{P\lambda_1\|\boldsymbol{\phi}^H\mathbf{u}_1\|^2}{BN_0}\label{eq:optimalsnr}
        \end{equation}
        where $B$ is the bandwidth.
        The $\mathbf{u}_1$  and $\lambda_1$ determine the performance of the system and depends on the considered channel conditions. In the following, we will obtain them under different ssRIS-UE channel conditions. 

    \subsection{Data rate under different channel conditions}

        While the ssRIS must be positioned within the BS's LOS region to meet self-sustainability requirements due to its passive nature, the ssRIS-UE channel need not be LOS. Thus, the data rate shall also be discussed separately based on the channel conditions. 
        
        \subsubsection{LOS ssRIS-UE channel}
            Under the LOS component-dominated assumption, the channel response from ssRIS to UE is
            \begin{equation}
                \mathbf{g} = \sqrt{\rho(\theta,d_\text{RD})}e^{-j\varphi_\text{RD}}\mathbf{a}_M(\theta)
            \end{equation}
            where $\varphi_\text{RD}$ is the reference phase shift and $\theta$ is the azimuth angle of the UE with respect to the ssRIS. 
            
            Following similar analysis as for $\mathbf{G}$, we can show that $\mathbf{V} = \operatorname{diag}(\mathbf{g})\mathbf{G}^H\mathbf{G} \operatorname{diag}(\mathbf{g})^H$ yields the dominant eigenvalue and eigenvector as
            \begin{align}
                \mathbf{u}_1=\frac{\overline{\mathbf{a}}_{M}}{\sqrt{\rho(\theta,d_\text{RD})M}},\quad
                 \lambda_1 = \rho_0 NM
            \end{align}
            where $\overline{\mathbf{a}}_{M}=\operatorname{diag}(\mathbf{g})\mathbf{a}_M^*(\psi)$ and $\rho_0 \triangleq \rho(\psi,d_\text{SR})\rho(\theta,d_\text{RD})$. 
            
            This leads to the optimal beamforming vectors $\boldsymbol{\phi}=\sqrt{M}\mathbf{u}_1$ and $\mathbf{w}= \frac{\mathbf{a}_N(\psi)}{\sqrt{N}}$, yielding the data rate
            \begin{equation}
                R = B\log_2\bigg(1+\underbrace{P\rho_0/(BN_0)}_{\triangleq \Gamma_0}\cdot NM^2\bigg)
            \end{equation}
            and optimal received power for energy harvesting
            \begin{equation}
                P_\text{Rc} = PNM\rho(\psi,d_\text{SR}).
            \end{equation}

        \subsubsection{NLOS ssRIS-UE channel}
            In the NLOS case, the propagation between ssRIS and UE is considered to go through arbitrary fading as
            \begin{equation}
                \mathbf{g} = \sqrt{\rho(\theta,d_\text{RD})}\tilde{\mathbf{g}}
            \end{equation}
            where $\tilde{\mathbf{g}}\in\mathbb{C}^{M}$ is the fading vector. Then $\mathbf{V}$ is expressed as
            \begin{align}
                \mathbf{V} &= \rho(\theta,d_\text{RD})\operatorname{diag}(\tilde{\mathbf{g}})\mathbf{G}^H\mathbf{G}\operatorname{diag}(\tilde{\mathbf{g}})^H\notag\\
                & = \rho_0N \Vert \tilde{\mathbf{a}}_M\Vert^2\frac{\tilde{\mathbf{a}}_M}{\Vert \tilde{\mathbf{a}}_M\Vert}\frac{\tilde{\mathbf{a}}_M^H}{\Vert \tilde{\mathbf{a}}_M\Vert}
            \end{align}
            where the introduced auxiliary vector $\tilde{\mathbf{a}}_{M}=\operatorname{diag}(\tilde{\mathbf{g}})\mathbf{a}_M^*(\psi)\in\mathbb{C}^{M}$.
            Given that, the eigenvalue and eigenvector can be obtained correspondingly as
            \begin{equation}
                \mathbf{u}_1=\frac{\tilde{\mathbf{a}}_M}{\Vert \tilde{\mathbf{a}}_M\Vert}, \quad \lambda_1= \rho_0 N\sum_{m=1}^{M}|\tilde{g}_m|^2,\label{eq:12} 
            \end{equation}
            where $\tilde{g}_m$ is the $m$-th entry of $\bar{\mathbf{g}}$. Then referring to~\eqref{eq:optimalsnr}, the maximum SNR can be expressed as
            \begin{equation}
                \Gamma^{\star}=\Gamma_0 N\left(\sum_{m=1}^M|\tilde{a}_{M,m}|\right)^2
            \end{equation}
            where $\tilde{a}_{M,m}$ represents the $m$-th element in $\tilde{\mathbf{a}}_M$. Also by implementing~\eqref{eq:12}, the resulting beamformer is calculated as $\mathbf{w}= \frac{\mathbf{a}_N(\psi)}{\sqrt{N}}$, which leads to the maximum harvested power.

            Fast fading has a stronger impact in NLOS scenarios, but practical hardware limitations restrict the ssRIS from reconfiguring at the same rate as the channel fluctuations. Therefore, for the NLOS scenario, we evaluate the outage probability for SNR, defined as the probability that the SNR falls below the threshold $\gamma$. Without losing generality, uncorrelated-Rayleigh fading $\tilde{\mathbf{g}}\sim\mathcal{CN}(\mathbf{0},\mathbf{I}_M)$ is considered. As proposed in~\cite{kundu2020ris}, we can approximate SNR by using the central-limit-theorem (CLT). Based on the approximation, $Y = \sum_{i=1}^M|\tilde{a}_{M,i}|$ can be approximated as a truncated Gaussian random variable\footnote{Truncated Gaussian distribution is considered since $Y$ must be always nonnegative. The approximation is shown to be accurate for $M\geq 10$ in \cite{kundu2020ris}. } with the CDF 
            \begin{equation}
                F_Y(y) = \begin{cases}
                    1 - CQ((y-\mu_Y)/\sigma_Y), & y\geq 0,\\
                    0, & y<0,
                \end{cases}     
            \end{equation}
            where $\mu_Y = M\sqrt{\pi}/2$ and $\sigma_Y^2 = M(4-\pi)/4$. In addition, $C=1/Q(-\mu_Y/\sigma_Y)$ where $Q(\cdot)$ is the Gaussian Q-function. The SNR outage probability, $P_\text{out}(\gamma)$, is given as
            \begin{align}
                P_\text{out}(\gamma) &= \mathbb{P}\{\Gamma < \gamma\} =F_Y\left(\sqrt{\frac{\gamma}{N\Gamma_0}}\right).
                \label{eq:outage}
            \end{align}
            
    \subsection{Self-sustainability constraints under various HaR schemes}
    
        The HaR schemes determine the operational logic of the ssRIS. In~\cite{10348506}, HaR schemes including ES, TS, and power splitting (PS) are analyzed. For the assumed LOS channel between BS and ssRIS, ES and PS exhibit equivalent performance, narrowing our analysis to ES and TS. In both HaR schemes, the operational phases can be broadly divided into the harvesting phase and the reflecting phase. During the harvesting phase, all the incoming signals are fully absorbed. In the reflecting phase, signals that impinge on the ssRIS will be reflected losslessly, and power is spent on maintaining the extra phase shifts that are injected into the reflected signal. Self-sustainability in this work refers to the condition where the total harvested energy exceeds the energy consumed during the reflecting phase.
    
        \subsubsection{Element splitting}
    
            In the ES case, the harvesting and reflecting are performed simultaneously using different metasurface elements. $M_\text{Rf}$ and $M_\text{Hr}$ denote the number of elements used for reflecting and harvesting, respectively. With this scheme, the channel from the BS to the ssRIS is split into harvesting related channel $\mathbf{G}_\text{Hr}\in\mathbb{C}^{N\times M_\text{Hr}}$ and reflecting related channel $\mathbf{G}_\text{Rf}\in\mathbb{C}^{N\times M_\text{Rf}}$. The self-sustainability constraint of ES can be expressed as
            \begin{equation}
                \eta P \rho(\psi,d_\text{SR}) N M_\text{Hr} \geq M_\text{Rf}P_0 \label{eq:esss}
            \end{equation}
            where $P_0$ is the power consumption of an element that is used for reflection and $\eta$ is the energy harvesting efficiency.
            
        \subsubsection{Time splitting}
    
            In the TS case, the harvesting phase and reflecting phase happen orthogonally over time. During the harvesting phase, all elements will be connected to the harvesting circuit and perform energy harvesting, and during the reflecting phase, elements will be connected to the reflecting circuit which enables them to do the signal phase-shifting. We denote the time portion that is used for energy harvesting as $\tau$, and consequently, the time portion that is used for reflecting is $1-\tau$. The self-sustainability constraint of TS is expressed as
            \begin{equation} 
                \tau \eta P \rho(\psi,d_\text{SR}) N M  \geq (1-\tau)MP_0 \label{eq:tsss}
            \end{equation}
            from which, it can be noticed that the element number $M$ can be cancelled at both sides of the inequality indicating that the self-sustainability condition does not depend on the element number.        
    
\section{ssRIS-Assisted System Optimization}

    The feasibility of utilizing ssRIS is characterized by how element requirements scale with system conditions. To quantify this, we formulate optimization problems that minimize the number of elements required to achieve a target data rate $R_0$ while maintaining self-sustainability. The resulting minimum element count directly reflects feasibility, with a lower element count required under given conditions indicating higher feasibility, as the ssRIS can achieve performance targets with fewer resources. To this end, we formulate the element number optimization problem $\textbf{P1}$ for ES when the ssRIS-UE channel is LOS as
    \begin{subequations}
        \begin{align}
            &\textbf{P1: }\underset{M_\text{Hr},M_\text{Rf}}{\text{minimize}} \quad M_\text{Hr}+M_\text{Rf}\\
            &\text{s.t.}\hspace{4mm} \eqref{eq:esss},\notag\\
            &\hspace{8mm} B\log_2\left(1+\Gamma_0NM^2_{\text{Rf}}\right)\geq R_0.\label{eq:esdata}
        \end{align}
    \end{subequations}
    where constraint~\eqref{eq:esss} guarantees the self-sustainability of the ssRIS, constraint~\eqref{eq:esdata} guarantees that the resulting data rate of the UE is higher than the threshold. As the total element requirement is not upper-bounded, the optimality can always be achieved. The optimal $M_\text{Hr}$ and $M_\text{Rf}$ can be obtained analytically as the equality is achieved for~\eqref{eq:esss} and~\eqref{eq:esdata}. The optimal number of elements can be calculated as
    \begin{align}
        M_\text{Rf}^{\star}&=\sqrt{\left(2^{\frac{R_0}{B}}-1\right)\frac{1}{N\Gamma_0}}\\
        M_\text{Hr}^{\star}&=\underbrace{P_0/(\eta PN\rho(\psi,d_\text{SR}))}_{\triangleq \alpha}\sqrt{\left(2^{\frac{R_0}{B}}-1\right)\frac{1}{N\Gamma_0}}.
    \end{align} 
    where the constant term $\alpha$ reflects the harvesting condition. Smaller $\alpha$ corresponding to favorable harvesting conditions with higher transmit power, smaller propagation attenuation $\rho(\psi,d_\text{SR})$, or more antennas $N$.

    The optimization problem for TS when the ssRIS-UE channel is LOS is denoted as $\textbf{P2}$, and is formulated as
    \begin{subequations}
        \begin{align}
        &\textbf{P2: }\underset{M, \tau}{\text{minimize}} \quad M\\
        &\text{s.t.}\hspace{4mm}\eqref{eq:tsss},\notag\\
        &\hspace{8mm} (1-\tau)B\log_2\left(1+\Gamma_0NM^2\right)\geq R_0.
    \end{align}
    \end{subequations}
    With a similar approach as applied to the ES case, we can solve the problem analytically, and the optimality is given as
    \begin{align}
        &\tau^{\star}=\frac{P_0}{\eta PN\rho(\psi,d_\text{SR})+P_0}\label{eq:opttau}=\frac{\alpha}{1+\alpha}\\
        &M^{\star}=\sqrt{\left(2^{\frac{R_0}{(1-\tau^{\star})B}}-1\right)\frac{1}{N\Gamma_0}}.\label{eq:optele}
    \end{align}
        
    When the ssRIS-UE channel experiences NLOS conditions, the affine constraint to the SNR is replaced by the outage probability constraint, requiring that $P_\text{out}(\gamma)$ remains below a tolerable margin $\varepsilon$. The optimization problem formulated for ES when the ssRIS-UE channel is NLOS is denoted as $\textbf{P3}$, which is given as 
    \begin{subequations}
        \begin{align}
        &\textbf{P3: }\underset{M_\text{Hr},M_\text{Rf}}{\text{minimize}} \quad M_\text{Rf}+M_\text{Hr}\\
        &\text{s.t.}\hspace{4mm} \eqref{eq:esss},\notag\\
        &\hspace{8mm} P_\text{out}(\gamma_0^\text{ES}) \leq \varepsilon\label{eq:esoutage},
        \end{align}
    \end{subequations}
    where $\gamma_0^\text{ES}$ is the corresponding SNR threshold for ES that is determined by $R_0$ as $\gamma_0^\text{ES} = 2^{\frac{R_0}{B}}-1$. With $\gamma_0^\text{ES}$ plugged into~\eqref{eq:outage}, constraint~\eqref{eq:esoutage} can be expanded as
    \begin{align}
        &\underbrace{(1-\varepsilon)Q\left(-\sqrt{\frac{M_\text{Rf}\pi}{4-\pi}}\right)}_{\triangleq f_1(M_\text{Rf})} \notag\\
        &\hspace{6mm}\leq \underbrace{Q\left(2\sqrt{\frac{\gamma_0^\text{ES}}{N\Gamma_0M_\text{Rf}(4-\pi)}}-\sqrt{\frac{M_\text{Rf}\pi}{4-\pi}}\right)}_{\triangleq f_2(M_\text{Rf})}.\label{eq:esult}
    \end{align}
    Noticing that $f_2(0)-f_1(0)<0$ and $f_2(\infty)-f_1(\infty)=\varepsilon>0$,  there exists at least one $M_\text{Rf}$ that satisfies the above inequality with equality. We let $\tilde{M}_\text{Rf}$ denote the minimum of these roots. This $\tilde{M}_\text{Rf}$ can be calculated numerically with the bisection method. The optimal value is then
    \begin{align}
        &M^{\star}_\text{Rf} = \tilde{M}_\text{Rf},\quad M_\text{Hr}^{\star} = \alpha M^{\star}_\text{Rf}.
    \end{align}
    
    Similarly, in the TS case, the optimization problem formulated is denoted as $\textbf{P4}$ and is given as
    \begin{subequations}
        \begin{align}
        &\textbf{P4: }\underset{M, \tau}{\text{minimize}} \quad M\\
        &\text{s.t.}\hspace{4mm} \eqref{eq:tsss}, \nonumber \\
        &\hspace{8mm} P_\text{out}(\gamma_0^\text{TS})\leq \varepsilon.\label{eq:tsoutage}
        \end{align}
    \end{subequations}
    We note that $\gamma_0^\text{TS}=2^\frac{R_0}{(1-\tau) B}-1$ is monotonically increasing with $\tau$, thus the outage probability is monotonically increasing with $\tau$. Thus, the minimum target SNR with $\tau$ from~\eqref{eq:opttau} is denoted $\tilde{\gamma}_0^\text{TS}$, and $\tilde{M}$ represents the number of elements required to meet the outage constraint. $\tilde{M}$ can also be calculated numerically by replacing $\gamma_0^\text{ES}$ and $M_\text{Rf}$ with $\tilde{\gamma}_0^\text{TS}$ and $\tilde{M}$ in \eqref{eq:esult}. The optimal $\tau^{\star}$ is obtained by~\eqref{eq:opttau}, and the optimal number of elements is obtained by finding the root that achieves the equality for~\eqref{eq:esult} after replacement via the bisection method.

\section{Numerical Analysis}

   From the previous analysis, the element requirement depends mainly on three factors: data rate requirement $R_0$, harvesting condition $\alpha$, and outage margin $\varepsilon$ (for NLOS scenarios). This section investigates ssRIS feasibility by examining how element requirements scale with each factor under different HaR schemes. While $\alpha$ can be reduced by increasing transmit power $P$, antenna count $N$, or improving hardware to lower $P_0$, we focus on varying $P$ as the most readily adjustable parameter in practical deployments. The impacts of $R_0$ and $\varepsilon$ are examined directly.

   We set the carrier frequency $f$ as 15\,GHz with a bandwidth $B$ of 50\,MHz. We consider $N = 128$, $\eta= 0.65$, and $P_0 =2\,\mu$W~\cite{10348506}. For the NLOS ssRIS-UE channel, we set the reference outage margin $\varepsilon$ to be 1$\%$ as a conservative empirical choice for high reliability. The length of the considered area, $d$, is set to be 50\,m. 

\begin{figure}[tb]
       \centering
       \includegraphics[width=0.75\linewidth]{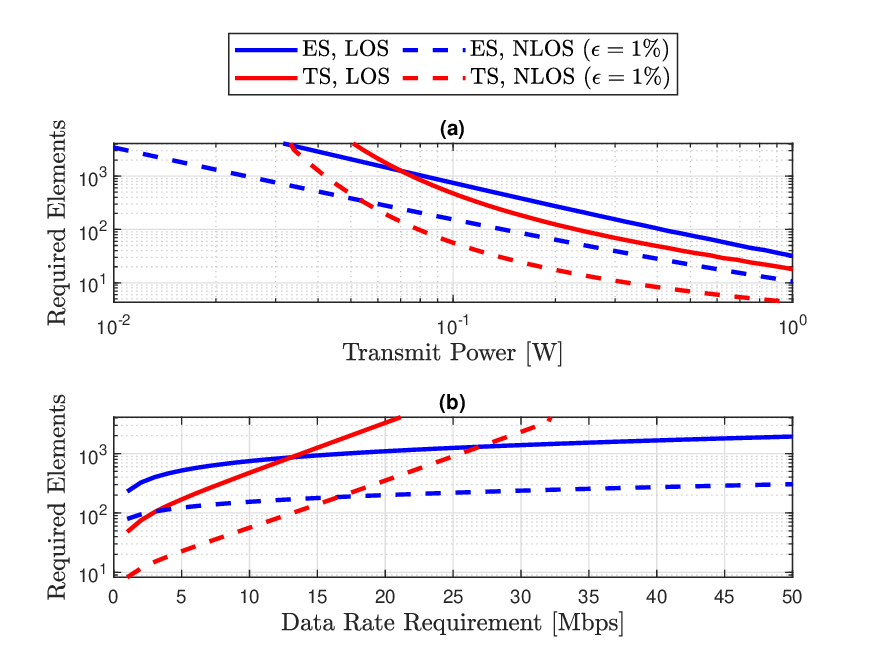}
       \vspace{-4mm}
       \caption{Element requirement across different system parameters. LOS and NLOS refer to the propagation conditions of the ssRIS-UE channel. (a) Impact of harvesting condition (via changing transmit power $P$) with $R_0 = 10$~Mbps; (b) Impact of data rate $R_0$ with $P = 0.1$~W.}
       \label{fig:PRVarying}
   \end{figure}

   As shown in Fig.~\ref{fig:PRVarying}, in the LOS case, TS achieves better element feasibility with fewer elements at high $P$ or low $R_0$. This advantage diminishes as $P$ decreases or $R_0$ increases, eventually leading to an intersection point where the advantage disappears. For analysis purposes, the total minimum element numbers required for ES and TS when the ssRIS-UE channel is LOS are expressed as
    $M_\text{ES} = (1+\alpha)\sqrt{\left(2^{\frac{R_0}{B}}-1\right)\frac{1}{N\Gamma_0}}$, and 
    $M_\text{TS} = \sqrt{\left(2^{\frac{R_0(1+\alpha)}{B}}-1\right)\frac{1}{N\Gamma_0}}$, respectively. As the transmit power is improved, $\alpha\rightarrow 0$, TS and ES behave similarly in terms of the element requirement. However, when the power harvesting condition is poor, $\alpha\rightarrow \infty$, the TS element requirements grow exponentially compared to ES's linear growth. Consequently, TS requires fewer elements when $\alpha$ is low, but significantly more when the harvesting condition gets worse. Similarly, as $R_0$ is directly scaled by $\alpha$, TS element requirements also grow faster as the increase of $R_0$ compared to the ES case. 

    The feasibility in the NLOS case is observed to follow a similar pattern as in the LOS case, yet with a more relaxed element requirement. The element requirement variations across different operating conditions in the NLOS case can be understood through the behavior of the outage probability constraint. Under favorable conditions (high transmit power and moderate data rate requirements), the root that reaches equality in~\eqref{eq:esult} lies around the central region of the Q-function, where both TS and ES schemes require similar and small element counts for communication purposes. However, ES additionally needs harvesting elements to support the reflecting operations, leading to higher total element requirements. Conversely, when operating conditions push the root to the tail region of the Q-function, the outage constraint becomes exponentially difficult to satisfy. Moreover, as $\gamma_0^{\text{TS}}$ becomes exponentially larger than $\gamma_0^{\text{ES}}$ for the same $R_0$, ES shows higher feasibility in this region.
    
\begin{figure}[tb]
       \centering
       \includegraphics[width=0.75\linewidth]{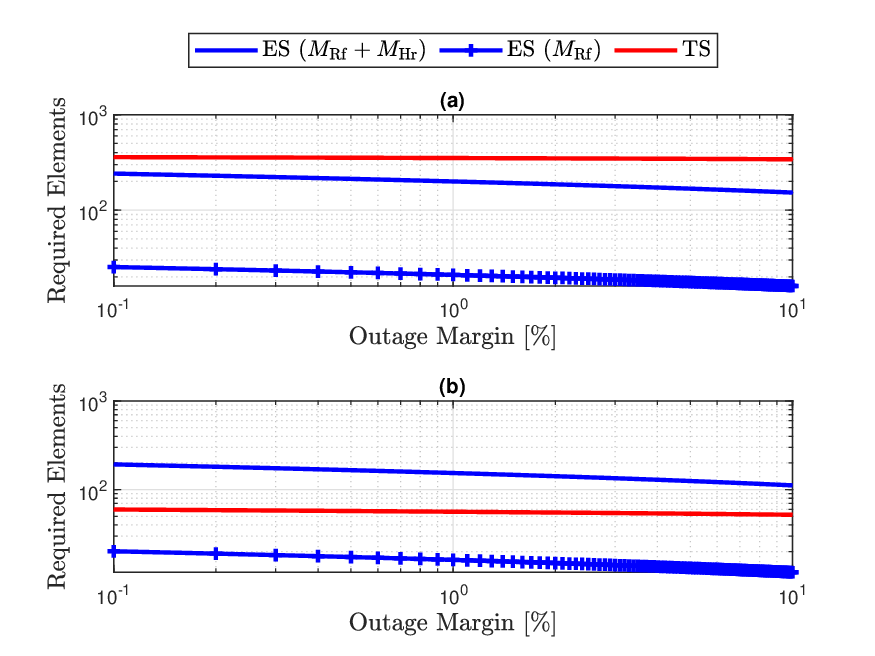}
       \vspace{-4mm}
       \caption{Element requirement across different outage margins under the NLOS ssRIS-UE channel. (a) $P=0.1$~W and $R_0 = 20$~Mbps; (b) $P=0.1$~W and $R_0=15$~Mbps.}
       \label{fig:outageVarying}
   \end{figure}

   To explicitly understand the impact of the outage margin, the element requirement is checked under the system condition near the intersection point shown in Fig.~\ref{fig:PRVarying}(a), and the corresponding results are illustrated in Fig.~\ref{fig:outageVarying}. As the outage constraint gets tightened with the decrease of $\varepsilon$, the root that achieves equality in~\eqref{eq:esult} will be pushed to the tail region to make the arguments of the Q-function on both sides have a smaller gap. The observations from Fig.~\ref{fig:outageVarying} verify this analysis, yet TS shows less sensitivity to the change in the outage margin. This can be explained by the fact that, for the same data rate requirement, more elements can be involved during the reflecting phase under the TS scheme, which brings an extra channel hardening gain.
   
\section{Conclusion}

    This study evaluates the feasibility of ssRIS deployment under ES and TS harvest-and-reflect schemes by analyzing how element requirements scale with harvesting conditions, data rate demands, and outage constraints. Both LOS and NLOS ssRIS-UE channel scenarios are examined for generality.

    The numerical results reveal complementary feasibility characteristics. TS exhibits better channel hardening, yet its element requirement exponentially scales with harvesting difficulties and data rate requirements, limiting its feasibility in challenging scenarios. This makes it ideal for indoor enhancing IoT and WiFi systems where stable connectivity is paramount and favorable harvesting conditions prevail. In contrast, ES demonstrates less sensitivity to harvesting conditions and data rate requirements. This makes ES more feasible for outdoor deployments where maintaining self-sustainability is more challenging.
    

\bibliographystyle{IEEEtran}
\bibliography{Main}
\end{document}